\documentstyle[aps,twocolumn,epsf]{revtex}
\def\be{\begin{equation}}
\def\ee{\end{equation}}
\def\q{{\bbox{q}}}
\def\r{{\bbox{r}}}
\def\lan{\langle}
\def\ran{\rangle}
\begin{document}
\title{Glass transition in the quenched and annealed version of the frustrated
lattice gas model} 
\author{Annalisa Fierro$^{1,2}$, Antonio de Candia$^{1,2}$ and
        Antonio Coniglio$^{1,2}$}
\address{$^1$Dipartimento di Scienze Fisiche,\\
         Complesso Universitario di Monte Sant'Angelo,
         Via Cintia, I-80126 Napoli, Italy}
\address{$^2$INFM, Sezione di Napoli, Napoli, Italy}
%
%
\maketitle
\begin{abstract}
In this paper we study the $3d$ frustrated lattice gas model in the 
annealed version, where the disorder is allowed to evolve in time with a 
suitable kinetic constraint. 
Although the model does not exhibit any thermodynamic transition it shows a
diverging peak at some characteristic time in the dynamical non-linear
susceptibility, similar to the results on the $p$-spin model in mean
field and Lennard-Jones mixture recently found by Donati {\em et al.}   
[{\bf cond-mat/9905433}]. Comparing these 
results to those obtained in the model with quenched interactions, we 
conclude that the critical behavior of the dynamical susceptibility is
reminiscent of the thermodynamic transition present in the quenched model, and 
signaled by the divergence of the static non-linear susceptibility, suggesting
therefore a similar mechanism also in supercooled glass-forming liquids.  
\end{abstract}
%
%
\section{Introduction}
The study of glass forming systems and spin glasses has shown that
these systems present a similar complex dynamical behavior. 
In both cases the relaxation time increases dramatically when the
temperature is lowered; furthermore
at temperatures lower than some temperature $T^*$,
the relaxation functions are well fitted at long times
by a stretched exponential function:
\be
f(t)=f_0\, \exp\left\{-\left({t\over\tau_\alpha}\right)^\beta\right\},
\qquad 0<\beta<1.
\label{stretch}
\ee
This similarity was further stressed by the observation that
the dynamical equations of a class of mean field spin glass models,
called $p$-spin glasses \cite{pspin}, are precisely equal
to the mode coupling equations for supercooled liquids.
The $p$-spin glasses are a generalization of
the spin glass model, where spins interact via three or more body
interactions.

Despite these analogies the connection in finite dimension 
between glass forming systems and spin glasses is not completely clear.
As far as  the static properties are concerned, spin glasses undergo a 
thermodynamic transition at a well defined temperature $T_{SG}$,
where the non-linear susceptibility diverges (a similar behavior is found also 
in $p$-spin in finite dimension \cite{campellone}).
The class of systems that show a transition of this kind
contains systems with very different microscopic structures,
but with two essential common characteristics, namely
the presence of competitive interactions (frustration) and of
quenched disorder.  
On the other hand, glass formers are a class of systems
where disorder is not originated by some fixed external 
variables, but is ``self-generated'' by the positions and orientation 
of particles. Differently from spin glasses,
in glass forming liquids there is no sharp thermodynamic transition,
characterized by the divergence of a thermodynamical quantity
analogous to the non-linear susceptibility in spin glasses.
However Donati {\em et al.} \cite{franz} have 
recently introduced a time dependent non-linear susceptibility
both for spin models and for structural glasses: they have shown that in the
mean field spherical model (where the mode coupling equations are exact), 
and in  a Lennard-Jones mixture studied by molecular dynamics simulation
\cite{glotzer}, the
dynamical susceptibility exhibits a maximum at some characteristic time, and
that this maximum diverges as the dynamical temperature $T_D$ is approached
from above. One may wonder whether the
presence of this maximum is related somehow to the presence of
``quasi''-quenched disorder self-generated in the liquid, resembling the
divergence of the non-linear susceptibility present in spin systems with
quenched disorder. To shed some light on this problem in this paper we want to
compare the properties of the frustrated lattice gas model, which has been
recently introduced in the context of the glass transition
\cite{varenna}, in two cases: when the disorder is quenched and when
the disorder is self-generated. 

In the quenched case the model is a spin glass diluted with lattice gas 
variable, that being constituted by diffusing 
particles is suited to study quantities such as the diffusion coefficient or 
the density autocorrelation functions, that are usually important in the study 
of liquids. This model exhibits in mean field \cite{jef} properties closely 
related to those of $p$-spin models. In $3d$, at low enough temperature, 
numerical simulations \cite{decandia} show  a behavior of the diffusion 
coefficient very 
similar to that experimentally observed in fragile liquids \cite{angel}.
Moreover the model presents a continuous static transition where the 
fluctuations of the order parameter, which coincide with the non-linear 
susceptibility, diverge. This property is absent in the infinite dimension 
$p$-spin models \cite{franz} and in supercooled liquids
studied by molecular dynamics simulations, 
as Lennard-Jones binary mixtures \cite{glotzer}.

Here we show the results obtained by studying the $3d$ frustrated
lattice gas model in the annealed version, where the 
interactions are allowed to evolve with a kinetic constraint \cite{Phdthesis}.
We find that the dynamical behavior fits quite well the behavior predicted by
the mode coupling theory. It is also easy to show that the model does not 
present any thermodynamic transition, consequently there is no divergence in
the non-linear susceptibility. However the dynamical non-linear susceptibility
exhibits a maximum with a behavior similar to that found 
in the mean field $p$-spin model and in molecular dynamics
simulation of Lennard-Jones mixture. We compare this behavior with the
corresponding quantities calculated in the quenched version and 
conclude that the thermodynamic transition, present in the quenched model and
signaled by the divergence of the static susceptibility, manifests
itself in the annealed model in the 
critical behavior of the dynamical susceptibility. This behavior seems a 
consequence of the fact at short enough time
the interactions can be considered as quenched variables.
Since the annealed model shows a behavior reminiscent of supercooled glass
forming liquids, these results suggest that also in glass forming liquids the
behavior of time dependent non-linear susceptibility 
can be due to the presence of
slow degree of freedom which acts for short time as quenched variables.
Although the annealed lattice gas model does not show any thermodynamic
transition, we cannot exclude that this is due to the absence of significant
interactions: typically in a real system the dynamical constraint corresponds to
some kind of interactions; in our case the static of the model 
is instead described by a trivial Hamiltonian, while the complex dynamics is 
due to the kinetic constraint.

In Sect. \ref{mct} we recall briefly the main results of the mode coupling 
theory for supercooled liquids; in Sect. \ref{dnls} we introduce the dynamical
non-linear susceptibility as defined by Donati {\em et al.} Finally in Sect. 
\ref{model} we  present the frustrated lattice gas model and the dynamical
behavior observed by numerical simulations in the quenched 
(Sect. \ref{qversion}) and in the annealed version (Sect. \ref{aversion}).   

\section{The mode coupling theory}
\label{mct} 
In order to compare the dynamical behavior of the frustrated lattice gas
with the predictions of the mode coupling theory (MCT) for supercooled liquids
\cite{gotze} we recall briefly the results of this theory. 
The equations of motion of the normalized spatial
Fourier transform of the density autocorrelation functions, $\Phi_q(t)$,
are evaluated under suitable approximations, and a dynamical transition, 
considered as an idealization of the glass transition, is observed:
at high temperature the solutions $\Phi_q(t)$
vanish at long time (liquid phase); 
at temperatures below a certain critical value $T_{\text{MCT}}$
solutions with a nonzero long time limit $f_q$ (called Debye-Waller factor)
appear (glass phase).
This transition is due to the non linearity of the equations, and no
thermodynamic phase transition is present.

Let us introduce some important quantities: 
the exponent parameter $0.5\le\lambda<1$,
that is a constant depending only on the system, and
the separation parameter $\sigma$, that is proportional to $x-x_c$,
where $x$ is the external control parameter 
(density or temperature) and $x_c$ is the 
critical value ($\sigma$ is chosen positive in the glass phase).
Via the transcendental equation
\be
{\Gamma^2(1-a)\over\Gamma(1-2a)}={\Gamma^2(1+b)\over\Gamma(1+2b)}=\lambda,
\label{trasc}
\ee
the exponent $\lambda$ determines two exponents,
$0<a<0.5$ and $0<b\le 1$, that rule
the relaxation of the system near the critical point.

The MCT predicts that in the $\beta-$regime, near the dynamical transition, 
the correlators can be written as
\be
\Phi_q(t)=f_q^c+h_q c_\sigma g_{\pm}(t/t_\sigma),\qquad \mathrm{for}~ 
t_0\ll t\ll\tau_\alpha
\label{paw}
\ee
where $c_\sigma=\sqrt{|\sigma|}$ and $\pm$ refers respectively to
the glass and liquid phase. The exponent $a$ fixes the short time behavior,
$g_{\pm}(t/t_\sigma)=(t/t_\sigma)^{-a}$ for $t_0\ll t\ll t_\sigma$, 
while for $t_\sigma\ll t\ll\tau_\alpha$ one has a constant in the
glass phase $g_{+}(t/t_\sigma)=(1-\lambda)^{1/2}$, and the
so called von Schweidler
law in the liquid phase $g_{-}(t/t_\sigma)=-B(t/t_\sigma)^b$.
Here two time scales, diverging as the critical point is approached from above,
appear:
\begin{eqnarray}
&t_\sigma=t_0|\sigma|^{-\delta}, \qquad &\delta=\frac{1}{2a};\\
&\tau_\alpha=t_0 B^{-1/b}|\sigma|^{-\gamma}, \qquad &\gamma=
\frac{1}{2a}+\frac{1}{2b};
\label{diverg}
\end{eqnarray}
and $t_0$ is a microscopic time characteristic of molecules motion.

Furthermore, in the liquid phase, the theory predicts the following scaling law
for the $\alpha-$relaxation ($t\gg\tau_\alpha$):
\be
\Phi_q(t)=\tilde{\Phi}_q\left({t\over\tau_\alpha}\right);
\ee
where the master curve $\tilde{\Phi}_q(t/\tau_\alpha)$
is well fitted by a stretched
exponential of the form (\ref{stretch}), with $0<\beta<1$ depending on the
particular correlator, but not on the temperature. This
functional form usually fits the experiments as well.

In order to test the predictions of the MCT in the $\beta-$regime,
Gleim and Kob \cite{gleim} have introduced the following quantity:
\be
R_q(t)={\Phi_q(t)-\Phi_q(t')\over\Phi_q(t'')-\Phi_q(t')},
\label{glkob}
\ee 
where $t'$ and $t''$ are arbitrary times in the $\beta$-regime ($t'\ne t''$).
>From Eq. (\ref{paw}) we can see that, if 
$\Phi_q(t)$ is in agreement with the leading-order prediction of the theory,
then $R_q(t)$ must be independent on $q$ in the 
$\beta-$regime.

\section{The dynamical non-linear susceptibility}
\label{dnls} 
Donati {\em et al.} \cite{franz} have recently defined a
dynamical non-linear susceptibility both for spin models and for structural 
glasses. They have shown that in the mean field $p$-spin spherical models
(where the mode coupling equations are exact) there is a characteristic
time where the dynamical susceptibility has a maximum,
and that this maximum diverges as the dynamical temperature $T_D$
is approached from above.

The Hamiltonian of the $p$-spin model is
\be
H=\sum_{i_1<\cdots<i_p} J_{i_1\cdots i_p} S_{i_1}\cdots S_{i_p},
\ee
where $p\ge 3$, the couplings $J_{i_1\cdots i_p}$ are Gaussian
with zero mean and variance
$1/N^{p-1}$, and the spins are real variables, with the global constraint
$\sum_{i=1}^{N}S_i^2\equiv N$, where $N$ is the number of spins.
The dynamical non-linear susceptibility $\chi (t)$ is defined by 
\be
\chi (t)=\beta N(\lan q(t)^2\ran -\lan q(t)\ran^2),
\ee 
where $q(t)={1\over N}\sum_i S_i(t') S_i(t'+t)$
is the overlap between the states at times $t'$ and $t'+t$.
Solving the equation of motion for $\chi(t)$  at temperature higher than
$T_D$, they find that  $\chi(t)$ displays a maximum as a function of time,
$\chi(t^*)$, which is shifted to larger times $t^*$ as $T$ approaches
$T_D$ from above and increases as
a power law $\chi(t^*)\propto (T-T_D)^{-\alpha}$.

A similar behavior has also been found in 
molecular dynamics simulation performed for a Lennard-Jones mixture.
\section{The frustrated lattice gas model}
\label{model}
\subsection{The quenched model}
\label{qversion} 
Recently a lattice model, which has mean field properties closely related to
those of $p$-spin models, has been introduced \cite{varenna} in 
connection with the glass transition.
This model is a diluted spin glass, which, being constituted
by diffusing particles, is suited to study quantities like the diffusion
coefficient, or the density autocorrelation functions, that are usually
important in the study of liquids. The Hamiltonian of the model is:  
\be
-\beta H = J\sum_{\lan ij \ran}
(\epsilon_{ij}S_i S_j - 1)n_in_j +\mu \sum_i n_i,
\label{flg}
\ee
where $S_i=\pm 1$ are Ising spins, $n_i=0,1$ are occupation variables,
and $\epsilon_{ij}=\pm 1$ are quenched and disordered interactions.

This model reproduces the Ising spin glass in the limit
$\mu\rightarrow\infty$ (all sites occupied, $n_i\equiv 1$).
In the other limit, $J\rightarrow \infty$,
the model describes a frustrated lattice gas with properties
recalling those of a ``frustrated'' liquid.
In fact the first term of Hamiltonian (\ref{flg}) implies that two
nearest neighbor sites can be freely occupied only if their spin variables
satisfy the interaction, that is if $\epsilon_{ij}S_iS_j=1$, otherwise
they feel a strong repulsion.

To make the connection with a liquid, we note that the internal degree of 
freedom $S_i$ may represent for example internal orientation of a particle 
with non symmetric shape.
Two particles can be nearest neighbors only if the relative orientation
is appropriate, otherwise they have to move apart.
Since in a frustrated loop the spins cannot satisfy all interactions, in this 
model particle configurations in which a frustrated loop is fully occupied are 
not allowed.  The frustrated loops in the model are the same of the spin glass 
model and correspond in the liquid to those loops which, due to geometrical
hindrance, cannot be fully occupied by the particles.  

Another possible interpretation of the model is that $n_i$ is the occupation 
variable of the $i$-th cell and $S_i$ indicates the position of the center of
mass of the particle inside the cell. $S_i$ can in principle assume many
different values corresponding to the coordinates of the center of mass inside
the cell.
The interaction should take into account that not all pair of internal degree
of freedom in two neighbors cells are allowed.
In the frustrated lattice gas model for simplicity we drastically reduce the
internal degree of freedom to only two values ($S_i=\pm 1$) and introduce the
interaction $\epsilon_{ij}=\pm 1$ between two neighbor spin variables to mimic
the local disorder of the medium.
In the first version of the model we assume that the local disorder are
quenched.

In the case $J=\infty$ the model has a maximum density 
$\rho_{\text{max}}\simeq 0.68$.
It has been shown that there exists some density $\rho_c\simeq 0.62$,
where the system has a transition of the type of $3d$ $p$-spin model 
\cite{antonio}, with a divergence of the 
static non-linear susceptibility
\be
\chi_{\text{SG}}={1\over N}\sum_{ij} [\lan S_i n_i S_j n_j\ran^2]
\label{eq_chistat}
\ee
where the average $\lan\cdots\ran$ is over the Boltzmann measure,
while the average $[\cdots]$ is over the disorder configurations
$\{\epsilon_{ij}\}$.

\begin{figure}
\begin{center}
\mbox{\epsfysize=5cm\epsfbox{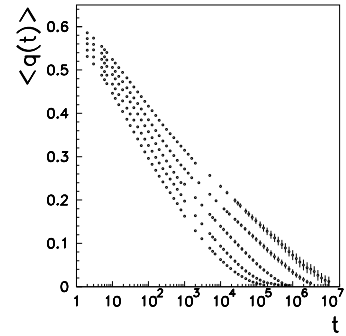}}
\end{center}
\caption{Relaxation functions of the self-overlap in the quenched model,
for a system of size $16^3$ and
densities $\rho=0.58$, $0.59$, $0.60$, $0.61$, $0.62$.}
\label{fig_qq}
\end{figure}

\begin{figure}
\begin{center}
\mbox{\epsfysize=5cm\epsfbox{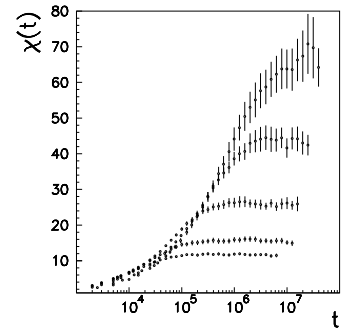}}
\end{center}
\caption{Dynamical susceptibility in the quenched model,
for the same system size and densities of Fig. \ref{fig_qq}.}
\label{fig_chiq}
\end{figure}

Here we show the results for the relaxation of the self-overlap,
which is defined as
\be
q(t)={1\over N}\sum_i  S_i(t')n_i(t')S_i(t'+t)n_i(t'+t),
\label{eq_q}
\ee
and for the dynamical susceptibility
\be
\chi(t)=N[\lan q(t)^2\ran-\lan q(t)\ran^2],
\label{eq_chi}
\ee
where the average $\lan\cdots\ran$ is done on the reference time $t'$.
In Fig. \ref{fig_qq} it is shown the relaxation functions $\lan q(t)\ran$
for a system of size $16^3$ for densities between $\rho=0.58$ and $0.62$.
Each curve is obtained averaging over a time interval for $t'$ of
$6\times 10^6$-$8\times 10^7$ Monte Carlo
steps, and finally averaging the results over $16$
realizations of the disorder.
Note that there is no sign of a two step relaxation. The long time tail
of the functions can be well fitted by a stretched exponential form,
with an exponent $\beta$ strongly dependent on the density,
which tends to very low values $\beta\simeq 0.2$ at high density.
In Fig. \ref{fig_chiq} it is shown the dynamical susceptibility
$\chi(t)$ for the same size and values of density
of Fig. \ref{fig_qq}. Note that $\chi(t)$
grows monotonically and has no maximum at finite time. The asymptotic
value $\chi(\infty)$ corresponds to the static susceptibility
(\ref{eq_chistat}), and therefore has a divergence at the density
$\rho_c\simeq 0.62$.
\subsection{The annealed model}
\label{aversion}
We have studied the frustrated lattice gas model (\ref{flg})
in the case where the interactions $\epsilon_{ij}=\pm 1$ are annealed. 
When we evaluate the partition function of the model, we must consider
in this case not only the $S_i$ and $n_i$, but also
the $\epsilon_{ij}$ as dynamical variables. Thus, summing over
the $\epsilon_{ij}$ and $S_i$ we obtain,
apart from an irrelevant factor,
$Z=\sum\limits_{\left\{n_i,S_i,\epsilon_{ij}\right\}}  e^{-\beta H}= 
\sum\limits_{\{n_i\}}e^{-\beta H_{\text{eff}}}$, where 
\be
-\beta H_{\text{eff}} = - K\sum_{\lan ij\ran} n_in_j +\mu \sum_i n_i, 
\label{eff}
\ee
and $K=-\ln\left[(1+e^{-2J})/2\right]$. Therefore the
static properties of the model are equal to those 
of a lattice gas with a repulsion between nearest neighbor particles, and
with no correlation between spins, $\lan S_i S_j\ran=\delta_{ij}$.
With the change of variables $n_i={1\over 2}(1+\sigma_i)$,
where $\sigma_i=\pm 1$ are Ising spins, this Hamiltonian can be 
written as the Hamiltonian of an antiferromagnetic Ising model
with an effective temperature $T_{\text{eff}}=4K^{-1}$, which is
always greater than the critical temperature
of the $3d$ antiferromagnetic Ising model $T_c\simeq 4.5$.
Therefore we can conclude that the model (\ref{eff}), and then also the model 
(\ref{flg}) with annealed interactions, does not present any thermodynamic 
transition. In the following we consider always the model with $J=\infty$.

We assume a dynamics for the variables $\epsilon_{ij}$
with a kinetic constraint, namely 
$\epsilon_{ij}$ can change its state only if 
the sites $i$ and $j$, and all their nearest neighbors, are empty;  
in this way the accessible states
to a given particle may change only if a wide enough region of the
system around it rearranges itself. We expect that, as the temperature
decreases, the disorder due to the local environment changes
so slowly that the interactions behave more and more as frozen playing the
role of ``self-induced quenched'' variables.
  
In order to generate an equilibrium configuration at a given density we
simulate the model without any dynamical constraint. In this case we can 
thermalize the system even at high density. Once an equilibrium configuration 
is obtained, we consider a diffusive dynamics for the particles while the 
interactions evolve with the kinetic constraint, as described before. In
conclusion the simulations are done in the following way: 
\begin{enumerate} 
\item[1)] one starts from an equilibrium configuration obtained at some density
$\rho$;
\item[2)] at each step of dynamics an interaction $\epsilon_{ij}$ is randomly 
chosen and
is changed if the sites $i$ and $j$, and all their nearest neighbors, are empty;
\item[3)]a particle (occupied site) on the lattice, one of the coordinate
directions, and a final state of the spin $S_i$ are randomly chosen;
\item[4)] one tries to
move the particle to the nearest neighbor site in the chosen direction.
The particle can move if two conditions are both satisfied. First, the
destination site must be empty. Second, the spins of the particles that are
nearest neighbors of the destination site, must satisfy the interaction with
the spin of the chosen particle.
If the particle cannot move in the chosen site, then
the move is rejected;
\item[5)] the clock advances one unit of time.        
\end{enumerate} 
During this dynamics we have evaluated relaxation functions and dynamic
non-linear susceptibility.
Note that as density grows, the relaxation time gets longer and longer, 
and gets longer than our observation time (which is between $10^7$ and 
$10^8$ for a system of size $16^3$) at a density approximately 
$\rho\simeq 0.63$.

\begin{figure}
\begin{center}
\mbox{\epsfysize=5cm\epsfbox{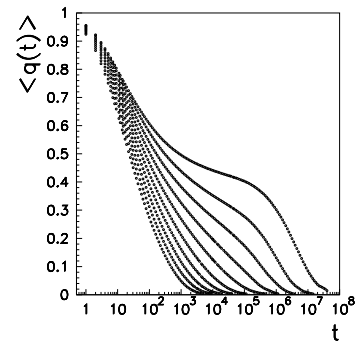}}
\end{center}
\caption{Relaxation functions of the self-overlap in the annealed model,
for a system of size $16^3$ and
densities $\rho=0.52$, $0.53$, $0.54$, $0.55$, $0.56$, $0.57$, $0.58$,
$0.59$, $0.60$, $0.61$.}
\label{fig_q}
\end{figure}

\begin{figure}
\begin{center}
\mbox{\epsfysize=5cm\epsfbox{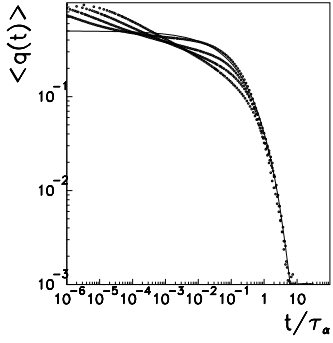}}
\end{center}
\caption{Time temperature superposition principle for the
relaxation functions of the self-over\-lap, for densities
$\rho=0.58$, $0.59$, $0.60$, $0.61$. The fitting function is a stretched
exponential with exponent $\beta=0.5$.}
\label{fig_ttsp}
\end{figure}

\begin{figure}
\begin{center}
\mbox{\epsfysize=5cm\epsfbox{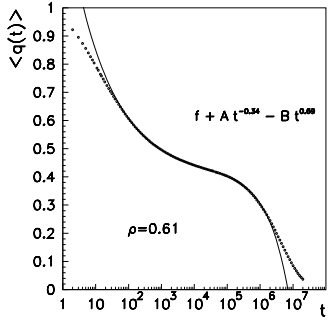}}
\end{center}
\caption{Fit of the intermediate time part of the relaxation function
of the self-overlap, for density $\rho=0.61$, with the fitting function
$f+At^{-a}-Bt^b$, where the fitting parameters are $f$, $A$, $B$ and
$\lambda$, and $a$ and $b$ are given by the relation (\ref{trasc}).}
\label{fig_mct}
\end{figure}

In Fig. \ref{fig_q} we show the relaxation functions of the self-overlap 
(\ref{eq_q}),
for a system of size $16^3$, for various densities between $\rho=0.52$
and $0.61$. 
Each curve is obtained averaging over a time interval for $t'$ of
$3\times 10^6$-$10^8$ Monte Carlo steps.
Observe that for high density the relaxation functions
clearly develop a two step relaxation, signaling the existence of two
well separated time scales in the system. We interpret the first
short time decay of the relaxation functions as due to the motion
of the particles in the frozen
environment, which on this time scales appear as quenched,
while the second decay is due to the evolution of environment, and 
final relaxation to equilibrium of the system.
The long time tail of the relaxation functions is well fitted by a 
stretched exponential form (\ref{stretch}), where the exponent $\beta$ 
depends very weakly from the temperature (it is constant within the errors)
and ranges between $\beta=0.4$ and $\beta=0.6$. In Fig. \ref{fig_ttsp}
we show the time-temperature superposition of the relaxation functions
of the overlap, for densities between $\rho=0.58$ and $\rho=0.61$.
We tried to fit the intermediate time part, corresponding to the
plateau, of the relaxation 
function of the overlap for density $\rho=0.61$, with the function 
predicted by the MCT (in a simplified form)
\be
\lan q(t)\ran=f+At^{-a}-Bt^b
\ee
where the fitting parameters are $f$, $A$, $B$ and $\lambda$, while
$a$ and $b$ are given by the relation (\ref{trasc}).
The result is shown in Fig. \ref{fig_mct}, where the full line is the fitting
curve with $a= 0.339\pm 0.002$ and $b=0.69\pm 0.01$.

\begin{figure}
\begin{center}
\mbox{\epsfysize=5cm\epsfbox{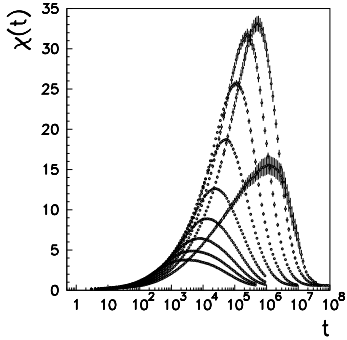}}
\end{center}
\caption{Dynamical susceptibility in the annealed model,
for the same system size and densities of Fig. \ref{fig_q}.}
\label{fig_chi}
\end{figure}

In Fig. \ref{fig_chi} we show the dynamical non-linear susceptibility
(\ref{eq_chi}) for the same size and values of the density
of Fig. \ref{fig_q}. It has the same behavior 
of the $p$-spin model in mean field and of the molecular dynamics simulation 
of the Lennard-Jones binary mixture \cite{franz},
namely a maximum $\chi(t^*)$ that seems to diverge
together with the time of the maximum $t^*$, when the density grows.
For the highest density , the maximum of $\chi(t)$ decreases, possibly
due to finite size effects, too short observation time,
or a change in the dynamics above 
some critical density. This fact is observed also in MD simulations
of Lennard-Jones liquids \cite{glotzer}.
We obtain that the maximum $\chi(t^*)$ as a function of the density
can be fitted quite well (taking out the last three points, where presumably
a rounding of the divergence takes place)
by the power law
$\chi(t^*)\propto (\rho_c-\rho)^{-\alpha}$, with $\rho_c=0.66\pm 0.01$
and $\alpha=3.6\pm 0.2$.
At very long times $\chi(t)$ decays to the equilibrium value, which is
simply $\chi(\infty)=\rho^2$. 
 
To make a more direct comparison with MCT, we have evaluated,
on a cubic lattice of size $8^3$,
the autocorrelation functions of the density fluctuations
\be
\Phi_{\q}(t)={\lan\delta\rho_{\q}(t'+t)\delta\rho _{\q}(t')
\ran\over\lan|\delta\rho _{\q}|^2\ran},
\ee
where the average $\lan\cdots\ran$ is performed on time $t'$, the density
fluctuation of wave number ${\q}$ is defined by
$\delta\rho_{\q}=\rho_{\q}-\lan\rho_{\q}\ran$, and
\be
\rho _{\q}(t)=\sum_{i=1}^{n} e^{-i{\q}\cdot{\r}_i(t)},
\ee
where ${\r}_i(t)$ is the position of the $i$-th particle at time $t$ and $n$
is the particle number.

Because of periodic conditions the allowable values of ${\q}$
on a cubic lattice are of the form
\be
{\q}={2\pi\over L}(n_x,n_y,n_z),
\ee
where $n_x,n_y,n_z=1\ldots{L\over 2}$ are integer values.

\begin{figure}
\begin{center}
\mbox{\epsfysize=5cm\epsfbox{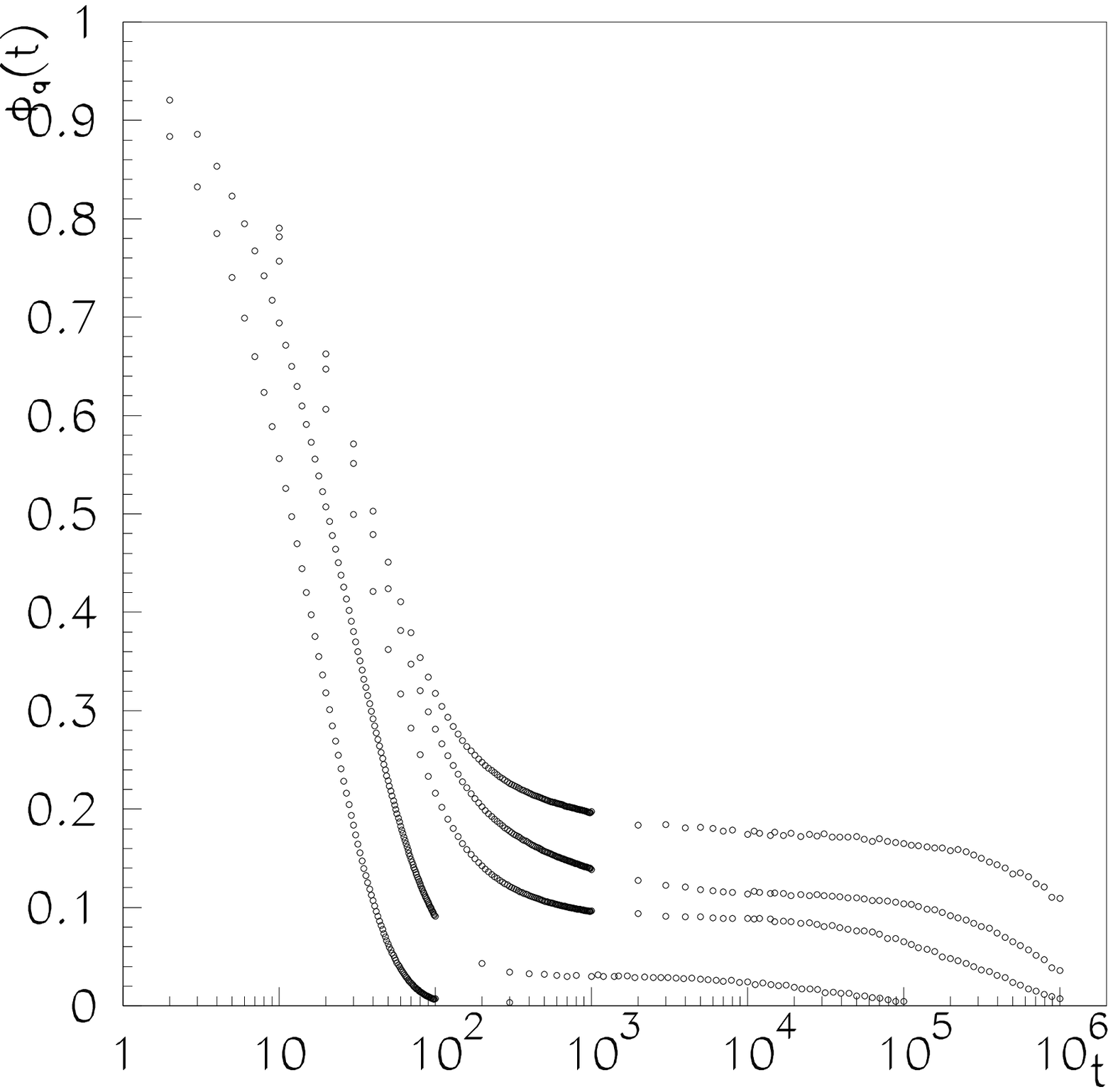}}
\end{center}
\caption{Correlation functions of the density fluctuations, $\Phi_q(t)$, for
${\q}=(\pi/4,0,0)$ at densities (from bottom to top)
$\rho=0.380$, $0.490$, $0.543$, 
$0.584$, $0.602$.} 
\label{fk1}
\end{figure}

\begin{figure}
\begin{center}
\mbox{\epsfysize=5cm\epsfbox{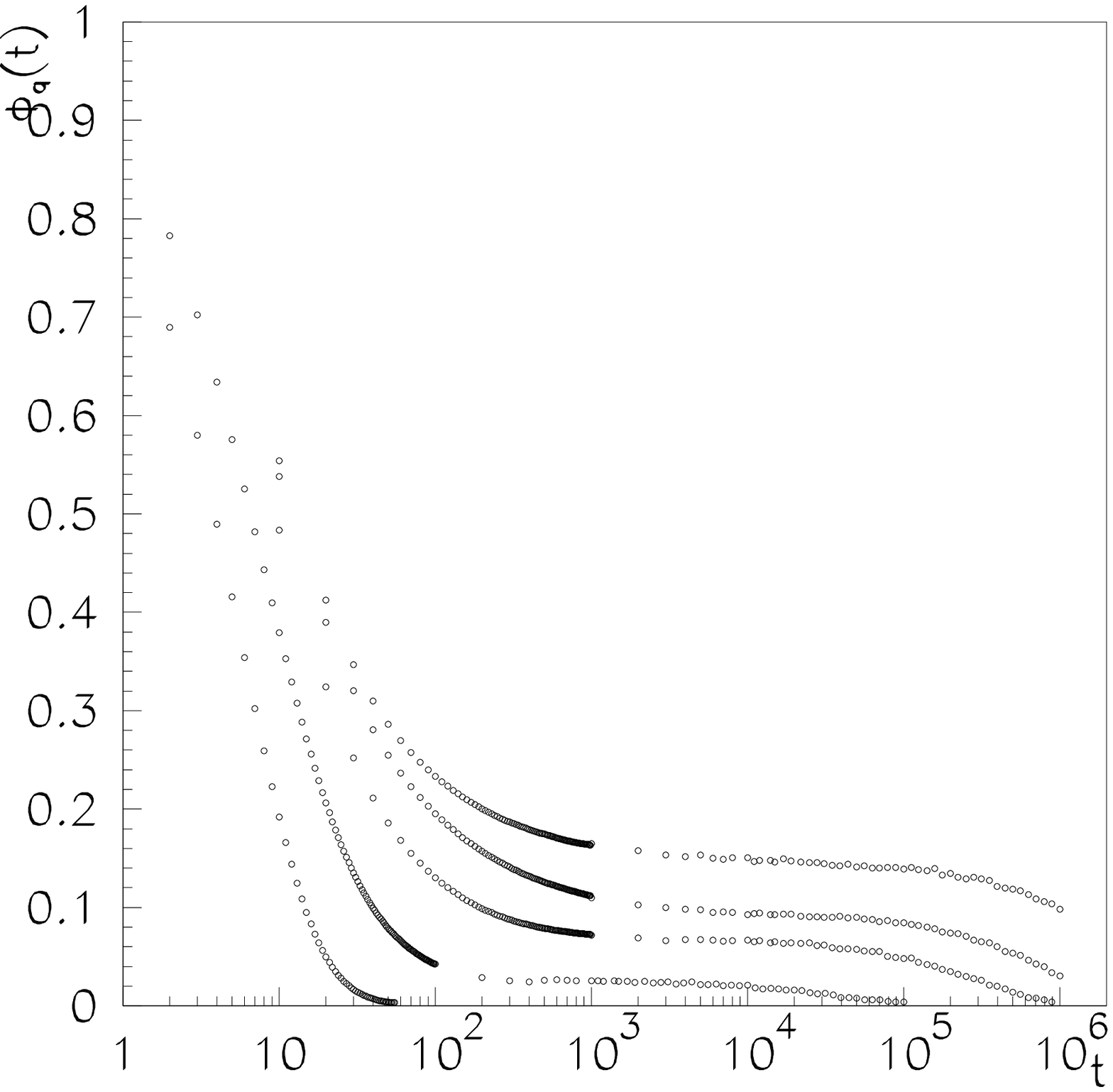}}
\end{center}
\caption{Correlation functions of the density fluctuations $\Phi_q(t)$ for
${\q}=(\pi/2,0,0)$ at densities (from bottom to top) 
$\rho=0.380$, $0.490$, $0.543$, 
$0.584$, $0.602$.}
\label{fk2}
\end{figure}

\begin{figure}
\begin{center}
\mbox{\epsfysize=5cm\epsfbox{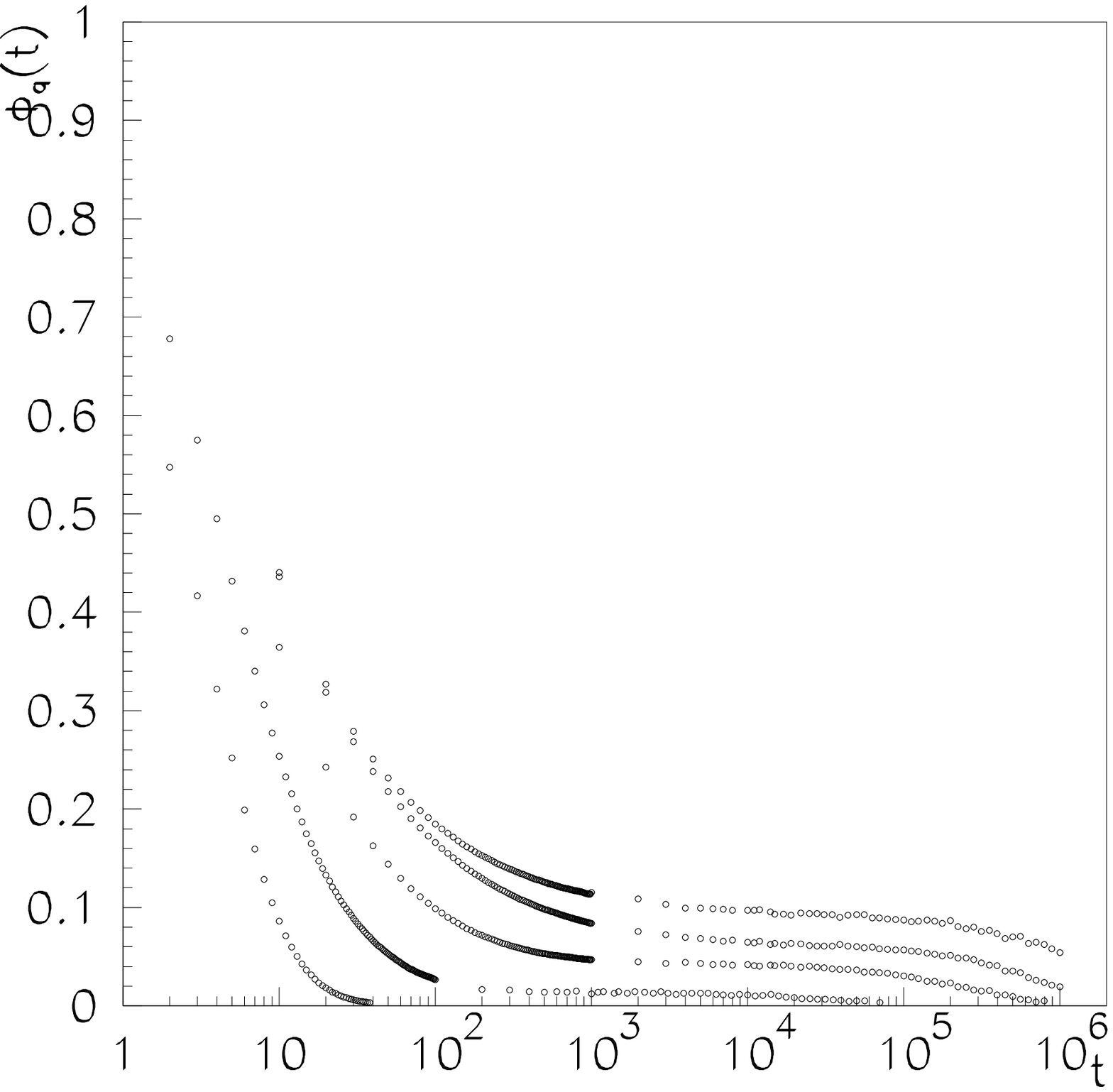}}
\end{center}
\caption{Correlation functions of the density fluctuations $\Phi_q(t)$ for  
${\q}=(\pi,0,0)$ at densities (from bottom to top)
$\rho=0.380$, $0.490$, $0.543$, 
$0.584$, $0.602$.}  
\label{fk3}
\end{figure}

In Fig. \ref{fk1}, \ref{fk2}, \ref{fk3} we show the results obtained at
densities between $\rho=0.380$ and $\rho=0.602$ respectively for 
${\bf q}=(\pi/4,0,0)$, $(\pi/2,0,0)$ and $(\pi,0,0)$. 
Each curve is obtained averaging over a time interval for $t'$ of $10^6$-$10^7$
steps and finally averaging the results obtained by $32-128$ 
different simulations.
As we can see in figures, at low density the
autocorrelation functions relax to zero with a one step decay; on the other
hand, as the density increases we can recognize the two-step decay
characteristic of glass forming systems.

\begin{figure}
\begin{center}
\mbox{\epsfysize=5cm\epsfbox{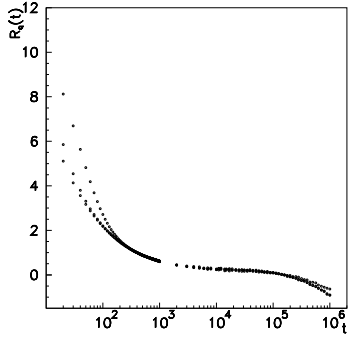}}
\end{center}
\caption{$R_q(t)$ for ${\q}=(\pi/4,0,0)$, $(\pi/2,0,0)$ and $(\pi,0,0)$ at
density $\rho=0.584$.} 
\label{kob}
\end{figure}

As we have said in Sect. \ref{mct}, if
$\Phi_{\q}(t)$ satisfies the prediction of the MCT, $R_q(t)$ 
(\ref{glkob}) must be independent on $q$ in the $\beta$-regime.
We have evaluated $R_q(t)$ (with $t''\simeq 400$ and $t'\simeq 1.6\cdot
10^5$) at some densities for all values of $q$ considered
here and we have obtained that $R_q(t)$ is independent on $q$ on a large time 
interval (see  Fig. \ref{kob}).
In agreement with this result we find that, after the initial
transient, the correlators $\Phi_q(t)$ are well fitted by a power law:
\begin{equation}
f_q+h_q\left(\frac{t}{t_0}\right)^{-a},
\label{paw1}
\end{equation}
and, in the intermediate time region, by a von Schweidler law:
\begin{equation}
f_q-h_q\left(\frac{t}{\tau}\right)^{b};
\label{von1}
\end{equation}
where the exponents $a$ and $b$ are independent on $q$.

\begin{figure}
\begin{center}
\mbox{\epsfysize=5cm\epsfbox{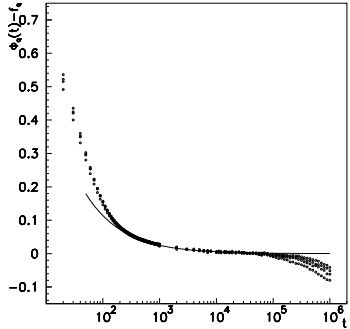}}
\end{center}
\caption{$\left[\Phi_q-f_q\right]$ as function of time $t$ for 
${\q}=(\pi/4,0,0)$ at
$\rho=0.548$, $0.573$,  $0.602$,  $0.615$. The full line is the power
law $(t/3.6)^{-0.66}$.}
\label{expa}
\end{figure}

\begin{figure}
\begin{center}
\mbox{\epsfysize=5cm\epsfbox{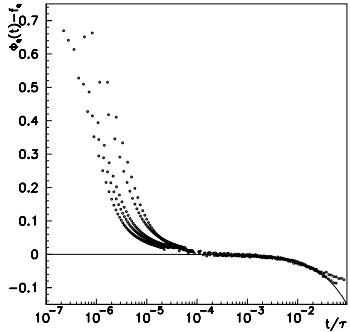}}
\end{center}
\caption{$\left[\Phi_q-f_q\right]$ as function of $t/\tau$ for 
${\q}=(\pi/4,0,0)$ at
$\rho=0.548$, $0.573$,  $0.602$,  $0.615$,  $0.625$. The full line is the von
Schweidler law $-(t/\tau)^{0.80}$.}
\label{expb}
\end{figure}

\begin{figure}
\begin{center}
\mbox{\epsfysize=5cm\epsfbox{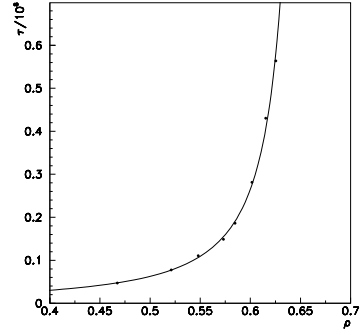}}
\end{center}
\caption{The relaxation time $\tau$ as function of density $\rho$. The full
line is the power law $(0.66-\rho)^{-1.5}$.}
\label{tau}
\end{figure}

In Fig. \ref{expa} and \ref{expb} we show $\left[\Phi_q(t)-f_q\right]$ as
function respectively of $t$ and $t/\tau$ for ${\bf q}=(\pi/4,0,0)$ at 
different values of densities. As we can see in figures the curves scale for all
values of density considered here and the data are in good agreement
respectively with the power law $(t/t_0)^a$ (with $a=0.66\pm 0.11$ and
$t_0=3.6\pm 3.0$) and the von Schweidler law $-(t/\tau)^b$ (with
$b=0.80\pm 0.13$). The relaxation time $\tau$, obtained as fitting parameter
from Eq. (\ref{von1}), is an increasing function of density, well represented by
a power law $(\rho_c-\rho)^{-\gamma}$, with $\rho_c=0.66\pm 0.03$ and 
$\gamma=1.5\pm 0.2$ (see Fig. \ref{tau}). 
However the values of the exponents $a$ and $b$ obtained in this case do not
satisfy the relation (\ref{trasc}) and are not in agreement with the exponents
obtained for $\lan q(t)\ran$.   

\begin{figure}
\begin{center}
\mbox{\epsfysize=5cm\epsfbox{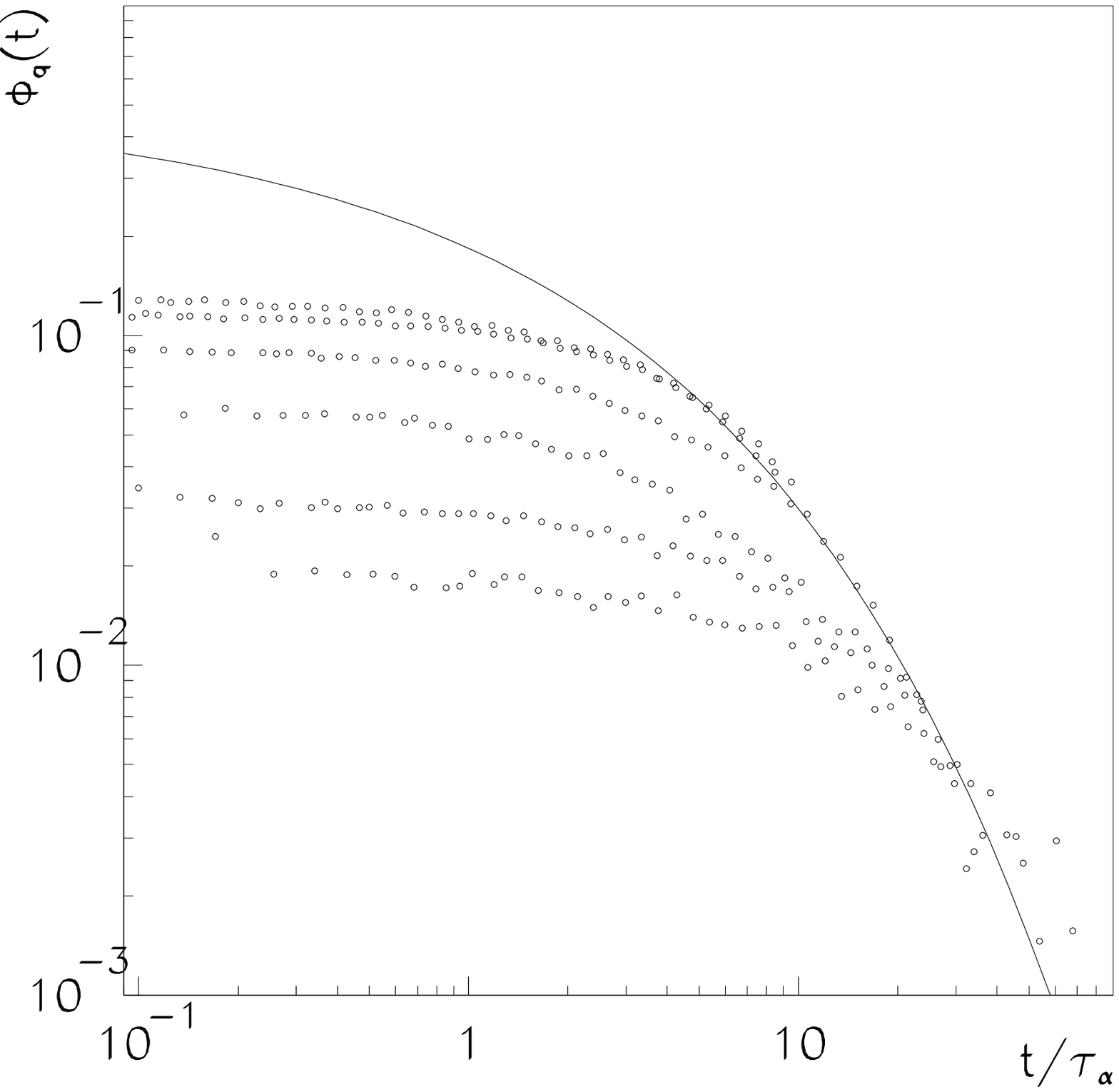}}
\end{center}
\caption{$\Phi_q$  as  function of $t/\tau_\alpha$ for ${\q}=(\pi/4,0,0)$ at
$\rho=0.467$, $0.490$, $0.521$, $0.543$, $0.573$, $0.584$. 
The full line is the stretched exponential function
$e^{-(t/\tau_\alpha)^{0.45}}$.}
\label{master}
\end{figure}

In Fig. \ref{master} we show the correlators
$\Phi_q(t)$ as functions of the rescaled times $\tilde{t}=t/\tau_\alpha$;
as we can see, the curves  coincide at large $\tilde{t}$, for all
values of density considered here, with a common master curve $\tilde{\Phi}_q$,
well fitted by a stretched exponential function (for ${\bf q}=(\pi/4,0,0)$ we
obtain $\beta\simeq 0.45$).  This result is consistent with the predictions
of the MCT concerning the $\alpha-$relaxation.
 
Finally as the density grows the equilibrium system smoothly
evolves towards an ``ordered'' state, analogous to the crystal state
(the system at high enough density can reach the equilibrium state only if 
frustration is reduced). As a
consequence of this fact, the similarity between the annealed model and a
supercooled glass forming liquid fails at high density.                         

\section{Conclusions} 
The frustrated lattice gas model in the quenched version presents a 
thermodynamic transition at a critical density, where the static non-linear 
susceptibility diverges.
The annealed model, which does not present any thermodynamic transition, 
consequently does not show any critical behavior of the static susceptibility; 
on the other hand we
observe an ``apparent'' divergence of the dynamical susceptibility at a value of
density $\rho_c$ (where the structural relaxation time also diverges).
We suggest that this behavior is due to the fact that at short enough time the
disorder can be considered as quenched. Moreover the similarity 
between the annealed model and the supercooled glass forming liquids suggests 
that also in these systems a similar mechanism may be responsible for the 
critical behavior of the dynamical susceptibility, and  one might speculate 
that if one could in special systems freeze some degree of freedom one could 
find a behavior similar to systems with quenched disorder.

\acknowledgements

This work was partially supported by the European TMR Network-Fractals under 
Contract No. FMR\-XCT\-980183 and from INFMPRA(HOP).

We acknowledge the allocation of computer resources from INFM Progetto Calcolo
Parallelo.


\begin{thebibliography}{999}
\bibitem{pspin}
A. Crisanti, H.J.Sommers, and H. Horner,  Z. Phys. B {\bf 87}, 341 (1992);\\
M. Mezard and G. Parisi, J. Phys I (France) {\bf 1}, 809 (1991);\\
L.F. Cugliandolo and J. Kurchan, Phys. Rev. Lett. {\bf 71}, 173 (1993);\\
S. Franz and M. Mezard, Europhys. Lett. {\bf 26}, 209 (1994).
%
\bibitem{campellone}
M. Campellone, B. Coluzzi, G. Parisi, Phys. Rev. B {\bf 58}, 12081 (1998);\\
M. Campellone, G. Parisi, P. Ranieri, Phys. Rev. B {\bf 59}, 1036 (1999).

\bibitem{franz} C. Donati, S. Franz, G. Parisi,  and S.C. Glotzer,
e-print  cond-mat/9905433.

\bibitem{glotzer} S.C.~Glotzer, V.N.~Novikov, and T.B.~Schr{\o}der,
cond-mat/9909113;\\
T.B.~Schr{\o}der, S.~Sastry, J.C.~Dyre, S.C.~Glotzer, cond-mat/9901271.

\bibitem{varenna} A. Coniglio, proceedings of the {\it International School
on the Physics of Complex Systems}, Varenna 1996;\\
M. Nicodemi and A. Coniglio, J. Phys A Lett. {\bf 30}, L187 (1996);\\
F. Ricci-Tersenghi, D. A. Stariolo and J. Arenzon, Phys. Rev. Lett. {\bf 84},
4473 (2000). 
%
\bibitem{jef} J. Arenzon, M. Nicodemi, and M. Sellitto,
Jour. de Phys. I (France) {\bf 6}, 1143 (1996).
%
\bibitem{decandia} A. de Candia, Ph. D. Thesis, Napoli (1998);\\
A. Coniglio, A. de Candia, A. Fierro and M. Nicodemi, J. Phys.: Condens.
Matter. {\bf 11}, A167 (1999).
%
\bibitem{angel} C. A. Angell, Science {\bf 267}, 1924 (1995).    
%
\bibitem{gotze}
W. Gotze, in {\em Liquids, Freezing and Glass Transition}, eds. J.P. Hansen,
D. Levesque, and Zinn-Justin, Elsevier (1991);
\\
T. Franosch, M. Fuchs, W. Gotze, M.R. Mayr and A.P. Singh,
Phys. Rev. E {\bf 55}, 7153 (1997);\\
M. Fuchs,W. Gotze and M.R. Mayr, Phys. Rev. E {\bf 58}, 3384 (1998). 
%
%
\bibitem{Phdthesis} A. Fierro, Ph. D. Thesis, Napoli (1999).
%
\bibitem{gleim} T. Gleim and W.Kob, cond-mat/9902003.  
%
\bibitem{antonio} A. de Candia and A. Coniglio, in preparation.
%
%
\end{thebibliography}
\end{document}